\documentclass[twocolumn,pra,showpacs,nofootinbib]{revtex4}
\usepackage[sort&compress]{natbib}
\usepackage{dcolumn,bm,graphicx,amsmath}

\begin{document}
\title{Application of the dual-kinetic-balance sets in the
relativistic many-body problem of atomic structure }

\author{ Kyle Beloy}
\affiliation{Physics Department, University of  Nevada, Reno, Nevada
89557}

\author{  Andrei Derevianko}
\email{andrei@unr.edu}
\homepage{http://wolfweb.unr.edu/homepage/andrei/tap.html}

\affiliation{Physics Department, University of  Nevada, Reno, Nevada
89557}
\affiliation{Laboratoire Aim\'{e} Cotton, B\^{a}t. 505, Campus d'Orsay, 91405 ORSAY Cedex France}
\date{\today}

\begin{abstract}
The dual-kinetic-balance (DKB) finite basis set method for solving
the Dirac equation for hydrogen-like ions [V. M. Shabaev {\em et
al.}, Phys. Rev. Lett. \textbf{93}, 130405 (2004)] is extended to
problems with a non-local spherically-symmetric Dirac-Hartree-Fock
potential. We implement the DKB method using B-spline basis sets and
compare its performance with the widely-employed approach of Notre
Dame (ND) group [W.R. Johnson and J. Sapirstein, Phys. Rev. Lett.
\textbf{57}, 1126 (1986)]. We compare the performance of the ND and
DKB methods by computing various  properties of Cs atom: energies,
hyperfine integrals, the parity-non-conserving amplitude of the
$6s_{1/2}-7s_{1/2}$ transition, and the second-order many-body
correction to the removal energy of the valence electrons. We find
that for a comparable size of the basis set the accuracy of both
methods is similar for matrix elements accumulated far from the
nuclear region. However, for atomic properties determined by small
distances, the DKB method outperforms the ND approach. In addition,
we present a strategy for optimizing the size  of the basis sets by
choosing progressively smaller number of basis functions for
increasingly higher partial waves. This strategy exploits
suppression of contributions of high partial waves to typical
many-body correlation corrections.
\end{abstract}

\maketitle

\section{Introduction}
Applications of perturbation theory in quantum mechanics require
summations over a complete set of states of the lowest-order
Hamiltonian. Usually, the relevant spectrum is innumerable. In
practical applications such eigenspectra are often modeled using
finite basis sets, chosen to be numerically complete. Since the sets
are finite, the otherwise infinite summations become amendable to
numerical evaluations.

The use of a finite basis set composed of piecewise polynomials,
so-called B-splines~\cite{deB01},  has proven to be particularly
advantageous in atomic physics and quantum chemistry
applications~\cite{BacCorDec01}. In this approach, an atom is placed
in a sufficiently large cavity and the atomic wavefunctions are
expanded in terms of the underlying B-spline set. Further, the
variational Galerkin method is invoked and the solution of the
resulting matrix eigenvalue problem produces a quasi-spectrum for
the atom. In non-relativistic calculations, the lowest-energy
orbitals  of the resulting basis set closely agree with those of the
unperturbed atom, and calculations of various properties of the
low-lying states can be carried out. In particular, one could
generate single-particle orbitals in some suitable lowest-order
approximation, and use the resulting basis set in applications of
many-body perturbation theory (MBPT) .

Application of the outlined approach to the {\em relativistic}
problems brings in a complication-- the appearance in the atomic
quasi-spectrum of non-physical ``spurious'' states. These states
appear in the solution of the single-particle radial Dirac equation
for $\kappa>0$ angular symmetry, $j=\ell-1/2$ ( $p_{1/2}, d_{3/2},
\ldots$ orbitals). The spurious states rapidly oscillate and,
moreover, spoil the mapping of the generated quasi-spectrum onto the
low-energy orbitals of the atom. At the same time they are required
for keeping  the set complete. The problem of spurious states was
discussed in the literature in details, see e.g.,
Ref.~\cite{Gol85,QuiGraWil87,JohBluSap88,Iga06}, and several
solutions were proposed. In the pioneering applications of the
B-splines in relativistic many-body problem by the Notre Dame group,
\citet{JohBluSap88} added an artificial potential spike centered at
the origin to the Hamiltonian matrix. The overall effect was to
shift the spurious states to higher energies thus restoring the
low-energy mapping to the physical states.   We will refer to the
sets generated using this prescription as the Notre Dame (ND) sets.
Another solution was to use ``kinetically-balanced''
sets~\cite{QuiGraWil87}, which related the small and the large
components of the basis set functions via the Pauli approximation.
Recently, an extension of this method was proposed in
Ref.~\cite{ShaTupYer04}. Here, due to additional relations between
the small and large components, the negative (Dirac sea) and
positive energy spectra are treated in a symmetric fashion. To
emphasize this built-in symmetry, the authors refer to their method
as the ``dual-kinetic-balance'' (DKB) approach. In both methods (by
contrast to the ND prescription), the spurious states were shown to
be completely eliminated from the quasi-spectrum.

Motivated by  the success of the DKB method in computing properties
of hydrogen-like ions\cite{ShaTupYer04,Iga06}, here we investigate
the suitability of the DKB method in modeling the spectrum of the
(non-local) Dirac-Hartree-Fock (DHF) potential. We compare the
performance of the ND and DKB methods by computing various
properties of Cs atom: single-particle energies, hyperfine
integrals, and the parity-non-conserving amplitude of the
$6s_{1/2}-7s_{1/2}$ transition. We find that for properties
involving matrix elements accumulated near the nucleus, the DKB
method outperforms the ND approach. Otherwise, if the electronic
integrals are accumulated far from the nucleus, both methods produce
results of a similar quality.

In addition, we investigate a possibility of using varying number of
basis set functions for  different angular symmetries.  Summations
over intermediate states in expressions of perturbation theory are
carried out both over angular quantum numbers $\kappa$ and for fixed
$\kappa$ over radial quantum numbers. Usually, as $|\kappa|$ (and
$\ell$) increases, the correlation corrections due to higher partial
waves become progressively smaller. Intuitively, one expects that
for higher partial waves it would be sufficient to  use smaller
radial basis sets of lesser quality. This would reduce storage
requirements for many-body calculations (for example, in
implementing coupled-cluster formalism ) and would speed up
numerical evaluations. While such an approach is common in quantum
chemistry, see, e.g., Ref.\cite{DavFel86}, the question of building
the optimal B-spline basis set was not addressed yet in relativistic
many-body calculations. We illustrate optimizing the basis sets by
computing the second-order energy correction for several states of
Cs.

This paper is organized as follows. First we recapitulate the
Galerkin-type approach to generating a finite-basis set
quasi-spectrum for the Dirac equation in Section~\ref{Sec:Setup}.
The variational method is invoked for relativistic action and the
problem is reduced to solving the generalized eigenvalue problem in
Section~\ref{Sec:Matrix}. The DHF potential is specified in
Section~\ref{Sec:Pot}. Further we specify ND and DKB sets in
Section~\ref{Sec:NDandDKB} and boundary conditions in
Section~\ref{Sec:SpuBnd}. A numerical analysis of Cs atom is
provided in Section~\ref{Sec:Numerics}. In Sections~\ref{Sec:EnHFI}
and~\ref{Sec:PNC} we compare the performance of the ND and the DKB
sets by computing single-particle energies, hyperfine integrals
(Section~\ref{Sec:EnHFI}), and parity-nonconserving amplitudes
(Section~\ref{Sec:PNC}) using both methods. The spurious states
arising from the ND method are examined in Section~\ref{Sec:Spur}.
In Section~\ref{Sec:E2} we consider second-order energy corrections
in the DKB method and discuss a strategy of optimizing the size of
the basis set.

\section{Problem setup}
\label{Sec:Setup}
We are interested in solving the eigenvalue equation $H_{D}u\left(
\mathbf{r}\right)  =\varepsilon~u\left(  \mathbf{r}\right)  $ for the Dirac
Hamiltonian, $H_{D}=c\mathbf{\alpha\cdot p}+\beta c^{2}+V_{\mathrm{nuc}%
}\left(  r\right)  +V_{\mathrm{DHF}}\left(  r\right)  $, where
$V_{\mathrm{nuc}}$ is the nuclear potential and $V_{\mathrm{DHF}}$
is the mean-field (Dirac-Hartree-Fock) potential. $V_{\mathrm{DHF}}$
is in general a non-local potential. Assuming that both potentials
are central one may exploit the rotational invariance to
parameterize the solutions as
\begin{equation}
u_{n\kappa}(\mathbf{r})=\frac{1}{r}\left(
\begin{array}
[c]{c}%
iP_{n\kappa}(r)\ \Omega_{\kappa m}(\mathbf{\hat{r}})\\
Q_{n\kappa}(r)\ \Omega_{-\kappa m}(\mathbf{\hat{r}})
\end{array}
\right)  ,\label{Eq:BiSpinorSpher}%
\end{equation}
with $\Omega_{\kappa m}(\mathbf{\hat{r}})$ being the spherical spinors. The
solutions depend on the radial quantum number $n$ and the angular quantum
number  $\kappa=(l-j)\left(
2j+1\right)  $. The large, $P_{n\kappa}$, and small, $Q_{n\kappa}$, radial
components satisfy the conventional set of radial Dirac equations
\begin{widetext}
\begin{align*}
\left(  V_{\mathrm{nuc}}\left(  r\right)  +V_{\mathrm{DHF}}\left(
r\right)
+c^{2}\right)  P_{n\kappa}\left(  r\right)  +c\left(  \frac{d}{dr}%
-\frac{\kappa}{r}\right)  Q_{n\kappa}\left(  r\right)   &
=\varepsilon
_{n\kappa}P_{n\kappa}\left(  r\right)  ,\\
-c\left(  \frac{d}{dr}+\frac{\kappa}{r}\right)  P_{n\kappa}\left(
r\right) +\left(  V_{\mathrm{nuc}}\left(  r\right)
+V_{\mathrm{DHF}}\left(  r\right) -c^{2}\right)  Q_{n\kappa}\left(
r\right)   &  =\varepsilon_{n\kappa }Q_{n\kappa}\left(  r\right)  .
\end{align*}
These radial equations may  be derived by seeking an extremum of the
following action~\cite{JohBluSap88}
\begin{eqnarray}
S &  = & \frac{1}{2}c\int \left\{  P\left(  r\right)
\hat{O}_{-}Q\left( r\right)  -Q\left(  r\right)  \hat{O}_{+}P\left(
r\right)  \right\} dr+\frac{1}{2}\int \left(  P\left( r\right)
,Q\left(  r\right) \right)  \hat{V}_{\mathrm{DHF}}\left(
\begin{array}
[c]{c}%
P\left(  r\right)  \\
Q\left(  r\right)
\end{array}
\right)  dr \nonumber \\
& + &  \frac{1}{2}\int V_{\mathrm{nuc}}\left(  r\right)  \left(
P\left( r\right)  ^{2}+Q\left(  r\right)  ^{2}\right) dr-c^{2}\int
Q\left( r\right)  ^{2}dr \label{Eq:Action}
-\varepsilon\frac{1}{2}\int \left(  P\left(  r\right) ^{2}+Q\left(
r\right)  ^{2}\right)  dr+\Delta S^{\mathrm{bnd}}+\Delta
S^{\mathrm{spur}},
\end{eqnarray}
\end{widetext}
where the $\kappa$-dependent Pauli operators are defined as%
\[
\hat{O}_{\pm}^{\kappa}=\frac{d}{dr}\pm\frac{\kappa}{r}.
\]
Radial integrals here and below have implicit limits from $r=0$ to
$r=R$, where $R$ is the radius of the confining cavity. Boundary
conditions may be imposed by adding the term $\Delta
S^{\mathrm{bnd}}$ to the action. The term $\Delta S^{\mathrm{spur}}$
controls the appearance of the spurious states in the
quasi-spectrum. We will specify these two terms below.

\subsection{Reduction to the matrix form}
\label{Sec:Matrix}
We employ two finite basis sets $\left\{  l_{i}\left(  r\right)  \right\}  $
and $\left\{  s_{i}\left(  r\right)  \right\}  ,$ $i=\overline{1,2N}$, which,
since we are interested in solving the (angularly-decoupled) radial equations,
may depend on the angular quantum number $\kappa$. We may expand the large and
small components in terms of these bases
\begin{eqnarray}
P\left(  r\right)   &  = &\sum_{i=1}^{2N}p_{i}~l_{i}\left(  r\right)
,\label{Eq:Pexpansion}\\
Q\left(  r\right)   &  = & \sum_{i=1}^{2N}p_{i}~s_{i}\left(  r\right)  , \nonumber
\end{eqnarray}
the expansion coefficients being the same for both the large and the small
components. The above expansions are plugged into the action,
Eq.~(\ref{Eq:Action}), and, further, its extremum is sought by varying the
expansion coefficients. As a result one arrives at the following generalized
eigenvalue equation%
\begin{equation}
A\vec{p}=\varepsilon B\vec{p}, \label{Eq:MatrixEigenValProb}
\end{equation}
where $A$ and $B$ are $2N\times2N$ matrixes and $\vec{p}$ is the
vector of expansion coefficients in Eq.(\ref{Eq:Pexpansion}).  The matrix
elements $A$  are given by%
\begin{equation}
A_{ij}=D_{ij}+V_{ij}-2c^{2}S_{ij}+A_{ij}^{\mathrm{bnd}}+A_{ij}^{\mathrm{spur}%
} \, .
\end{equation}
The matrixes entering the definition of $A$ correspond to various pieces of
the radial  Dirac equations,
\begin{align}
D_{ij}&=   c\left(  \int l_{i}\left(  r\right)  \frac{d}{dr}%
s_{j}\left(  r\right)  dr-\int l_{i}\left(  r\right)  \left(
\frac{\kappa}{r}\right)  s_{j}\left(  r\right)  dr\right)\nonumber\\
&+c\left(  \int l_{j}\left(  r\right)  \frac{d}{dr}%
s_{i}\left(  r\right)  dr-\int l_{j}\left(  r\right)  \left(
\frac{\kappa}{r}\right)  s_{i}\left(  r\right)  dr\right),\label{Eq:Dij}\\
V_{ij} & =\int V_{\mathrm{nuc}}\left(  r\right)  \left[ l_{i}\left(
r\right)  l_{j}\left(  r\right)  +s_{i}\left(  r\right) s_{j}\left(
r\right)  \right]  dr%
\nonumber\\&
+\left(  V_{\mathrm{DHF}}\right)  _{ij},\\
S_{ij} &=\int s_{i}\left(  r\right)  s_{j}\left(  r\right)  dr,
\end{align}
with the matrix elements of the DHF potential, $\left(  V_{\mathrm{DHF}%
}\right)  _{ij}$,  given below. The terms $A_{ij}^{\mathrm{bnd}}$
and $A_{ij}^{\mathrm{spur}}$ arise from the boundary and
\textquotedblleft spurious state\textquotedblright\ corrections in
the action. Finally, the matrix $B$, $B_{ij}=\int\left[  l_{i}\left(
r\right)  l_{j}\left( r\right)  +s_{i}\left(  r\right)  s_{j}\left(
r\right)  \right]  dr,$ reflects the fact that the basis sets may be
non-orthonormal.

\subsection{ Potentials}
\label{Sec:Pot}
The nuclear potential $V_{\mathrm{nuc}}\left(  r\right)  $
is generated for a nucleus of a finite size. We employ the Fermi distribution
with the nuclear parameters taken from Ref.~\cite{JohSof85}. As
to the Dirac-Hartree-Fock potential,  we employ the frozen-core
approximation. In this method, the calculation is carried out in two stages.
First, the core orbitals are computed self-consistently. Second, based on the
precomputed core orbitals, the DHF potential is assembled for the valence
orbitals and the valence orbitals are determined. In the valence part of the
problem, the core orbitals are no longer adjusted. Explicitly, for a set of
the angular symmetry $\kappa$,%
\begin{align}
\left(  V_{\mathrm{DHF}}\right)  _{ij} &  =\int \left(
\begin{array}
[c]{cc}%
l_{i} & s_{i}%
\end{array}
\right)  \hat{V}_{\mathrm{DHF}}\left(
\begin{tabular}
[c]{l}%
$l_{j}$
\\
$s_{j}$%
\end{tabular}
 \right)
\nonumber\\
& =\left(  V_{\mathrm{DHF}}^{\mathrm{dir}}\right)
_{ij}+\left(  V_{\mathrm{DHF}}^{\mathrm{exc}}\right)  _{ij},\label{Eq:VDHF}\\
\left(  V_{\mathrm{DHF}}^{\mathrm{dir}}\right)  _{ij} &  =\sum_{a\in
\mathrm{core}}\left(  2j_{a}+1\right)  \nonumber\\
&\times\int v_{0}\left(  a,a,r\right) \left[  l_{i}\left(  r\right)
l_{j}\left(  r\right)  +s_{i}\left(  r\right)
s_{j}\left(  r\right)  \right]  dr,\\
\left(  V_{\mathrm{DHF}}^{\mathrm{exc}}\right)  _{ij} &  =-\sum_{a\in
\mathrm{core}}\sum_{L}\left(  2j_{a}+1\right)  \Lambda_{\kappa L\kappa_{a}}\nonumber\\
&\times\int v_{L}\left(  a,j,r\right)  \left[  l_{i}\left(  r\right)
P_{a}\left(  r\right)  +s_{i}\left(  r\right)  Q_{a}\left(  r\right)
\right]dr ,
\end{align}
with the conventionally defined multipolar contributions
\begin{align}
v_{L}\left(  b,a,r\right) =
\int \frac{r_{<}^{L}}{r_{>}^{L+1}%
}\left[  P_{a}\left(  r^{\prime}\right)  P_{b}\left(  r^{\prime}\right)
+Q_{a}\left(  r^{\prime}\right)  Q_{b}\left(  r^{\prime}\right)  \right]
dr^{\prime},
\end{align}
and the angular coefficient%
\begin{equation}
\Lambda_{\kappa_{a}L\kappa_{b}}=\left(
\begin{array}
[c]{ccc}%
j_{a} & j_{b} & L\\
-1/2 & 1/2 & 0
\end{array}
\right)  ^{2}.
\end{equation}

\subsection{ ND and DKB sets}
\label{Sec:NDandDKB} Now we specify the Notre Dame (ND) and the
dual-kinetic-balance (DKB) basis sets. Both operate in terms of
B-spline functions and first we recapitulate the relevant properties
of these splines. A set of $n$ B-splines of order $k$ is defined on
a supporting grid $\left\{  t_{i}\right\},i=\overline{1,n+k}  $.
Usually, the gridpoints are chosen as
\begin{align*}
t_{1} &  =t_{2}=\cdots=t_{k}=0,\\
t_{n} &  =t_{n+1}=\cdots=t_{n+k}=R.
\end{align*}
In our calculations the intermediate gridpoints are distributed
exponentially. B-spline number $i$ of order $k\,$, $B_{i}^{\left(
k\right)  }\left( r\right)  $,\ is a piecewise polynomial of degree
$k-1$ inside $t_{i}\leq r\ <t_{i+k}$. It vanishes outside this
interval. This property simplifies the evaluation of matrix elements
between the functions of the basis set. In addition, we will make
use of the fact that as $r\rightarrow0$, the first $k$ splines
behave as (all the
remaining splines are zero)%
\begin{equation}
B_{i\leq k}^{\left(  k\right)  }\propto r^{i-1},\label{Eq:BsplineAtr0}%
\end{equation}
and at $r=R$, all splines vanish except for the last spline, $B_{i=n}^{\left(
k\right)  }$.

The Notre Dame set is defined as%
\begin{align}
l_{i}^{\mathrm{ND}}\left(  r\right)   &  =\left\{
\begin{array}
[c]{cc}%
B_{i}^{\left(  k\right)  }\left(  r\right)   & 1\leq i\leq n\\
0 & n<i\leq2n
\end{array}
\right.  ,\label{Eq:NDset}\\
s_{i}^{\mathrm{ND}}\left(  r\right)   &  =\left\{
\begin{array}
[c]{cc}%
0 & 1\leq i\leq n\\
B_{i-n}^{\left(  k\right)  }\left(  r\right)   & n<i\leq2n
\end{array}
\right.  .\nonumber
\end{align}
It corresponds to an independent expansion of the large and small radial
components into the B-spline basis. The DKB set involves the Pauli operators
and enforces a \textquotedblleft kinetic balance\textquotedblright\ between
contributions to the components:%
\begin{align}
l_{i}^{\mathrm{DKB}}\left(  r\right)   &  =\left\{
\begin{array}
[c]{cc}%
-B_{i}^{\left(  k\right)  }\left(  r\right)   & 1\leq i\leq n\\
-\frac{1}{2c}\hat{O}_{-}^{\kappa}~B_{i-n}^{\left(  k\right)  }\left(
r\right)   & n<i\leq2n
\end{array}
\right.  ,\label{Eq:DKBset}\\
s_{i}^{\mathrm{DKB}}\left(  r\right)   &  =\left\{
\begin{array}
[c]{cc}%
\frac{1}{2c}\hat{O}_{+}^{\kappa}~B_{i}^{\left(  k\right)  }\left(  r\right)
 & 1\leq i\leq n\\
B_{i-n}^{\left(  k\right)  }\left(  r\right)   & n<i\leq2n
\end{array}
\right.  .\nonumber
\end{align}
Notice that, as discussed below, to satisfy the boundary conditions, we will
use only a subset of the entire DKB basis.

\subsection{Spurious states and boundary conditions }
\label{Sec:SpuBnd}
With the ND set, the spurious states are shifted away
to the high-energy end of the quasi-spectrum by adding the following
action~\cite{JohBluSap88} $\left(  \Delta S^{\mathrm{spur}}\right)  _{\mathrm{ND}}=\frac
{c}{2}P\left(  0\right)  ^{2}-\frac{c}{2}P\left(  0\right)  Q\left(  0\right)
$ for $\kappa<0$ and $\left(  \Delta S^{\mathrm{spur}}\right)  _{\mathrm{ND}%
}=c^{2}P\left(  0\right)  ^{2}-\frac{c}{2}P\left(  0\right)  Q\left(
0\right)  $ for $\kappa>0$. This correction may be seen as arising from an
artificial $\delta-$function-like potential centered at the origin.
Unfortunately, as shown below in numerical examples, this additional spike
perturbs the behavior of the orbitals near the nucleus. (As discussed below
$\left(  \Delta S^{\mathrm{spur}}\right)  _{\mathrm{ND}}$ also sets the boundary
conditions at $r=0.$) The DKB set does not have the spurious states at all, so
that  $\left(  \Delta S^{\mathrm{spur}}\right)_{\mathrm{DKB}}\equiv0$.

We need to specify boundary conditions at $r=0$
and at the cavity radius, $r=R$. We start by discussing the boundary
conditions at $r=0$. For a finite-size nucleus the radial components behave
as
\begin{align}
P_{n\kappa}  & \propto r^{l+1}~\mathrm{and}~Q_{n\kappa}\propto r^{l+2}%
~\mathrm{for~}\kappa<0~,\label{Eq:PQsmallr}\\
P_{n\kappa}  & \propto r^{l+1}~\mathrm{and}~Q_{n\kappa}\propto r^{l}%
~~~~\mathrm{for~}\kappa>0~.\nonumber
\end{align}
In the Notre Dame approach, the boundary conditions are imposed
variationally by adding the boundary terms to the action integral.
Varying $\Delta S^{\mathrm{spur}}$, Ref.\cite{JohBluSap88},
effectively reduces to  $P\left(  0\right)  =0$. In practice,
because of the variational nature of the ND constraint, the large
component, although being small, does not vanish at the origin, and
the limits, (\ref{Eq:PQsmallr}), are not satisfied. Alternatively,
\citet{FisPar93}  proposed to impose $P\left(  0\right)  =Q(0)=0$ by
discarding the first B-spline of the set (this is the only B-spline
that does not vanish at $r=0$). This is a \textquotedblleft
hard\textquotedblright\ constraint, since  the wavefunction,
Eq.(\ref{Eq:Pexpansion}), would vanish identically at the origin. In
our calculations,  because of our motivation in building the
smallest possible basis set,  we extend this scheme further. We
exploit the power-law behavior of the B-splines,
Eq.(\ref{Eq:BsplineAtr0}), and match it to the small-$r$ limits
(\ref{Eq:PQsmallr}). To satisfy the matching, we need to include the
B-splines starting with the sequential number ($i_{\min}$ must be
smaller than the order of the splines $k$).
\begin{equation}
i_{\min}= |\kappa|+1=\left\{
\begin{array}
[c]{cc}%
\ell+2, & \kappa<0\\
\ell+1, & \kappa>0
\end{array}
\right.  .
\label{Eq:imin}
\end{equation}
For $s_{1/2}\left(  \kappa=-1\right)  \,$\ and $p_{1/2}\left( \kappa
=+1\right)  ,\,i_{\min}=2,$ and this is equivalent to the boundary
condition of Ref.~\cite{FisPar93}. For higher partial waves,
however, an increasingly larger number of splines is discarded:
e.g., for $f_{7/2}\left(  \kappa=-4\right) ,~i_{\min}=5$. One should
notice that for a basis that includes partial waves \
$\ell\leq\ell_{\max}$, for a faithful representation of the
small-$r$ behavior in all the partial waves, one needs to require
the order of the splines to be at least of $k=\ell_{\max}+3$. In
particular, for  $k=7$, $l_{\max}=4$.

When the first B-spline of the set, $B_{i=1}^{\left(  k\right)
}\left(  r\right)$, is included in the basis (as in the ND
approach), there is another difficulty in the calculations: since
it's value does not vanish at $r=0$, the matrix elements $D_{1,n+1}$
and $D_{n+1,1}$, which contain matrix elements of $1/r$, are
infinite in absolute value~\cite{Iga06}. In practical calculations,
one uses Gaussian quadratures to evaluate this integral, so the
result of the integration is finite. Yet this introduces
arbitrariness in the ND calculations and may be a reason for a
relatively poor representation of the orbitals near the origin.

At the cavity radius,  to avoid overspecifying the boundary
conditions for the Dirac equation, the ND group used the boundary
condition $P\left(  R\right)  =Q\left(  R\right)  $. As with the
conditions at the origin, this relation was \textquotedblleft
encouraged\textquotedblright variationally. In our calculations
(similarly to Ref.~\cite{ShaTupYer04}) we use the \textquotedblleft
hard\textquotedblright\ condition $P\left(  R\right) =Q\left(
R\right)  =0$ by removing the last B-spline from the set.
Examination of the resulting orbitals reveals that the wavefunctions
acquire a non-physical inflection towards the end of the supporting
grid, while the ND orbitals behave properly. Further numerical
experimentation, however, shows that the inflection does not degrade
the numerical quality at least for the atomic properties of the
low-lying bound states of interest.  At the same time, throwing away
the last B-spline reduces the number of  basis functions and leads
to a more compact set.

To summarize, we will use the DKB basis set that includes B-splines
with sequential numbers $i_\mathrm{min} (\kappa)$,
Eq.~(\ref{Eq:imin}) to $i_\mathrm{max}=n-1$. We will simply refer to
this choice as the DKB basis. When we refer to $N=40$ DKB functions
for a given partial wave, it would imply larger underlying B-spline
set, e.g., for $s$-waves the total number of B-splines would be $n =
42$. For the ND basis $n=N$, as all B-splines participate in the
expansion.

In the remainder of this paper we  present results of numerical
analysis for  $^{133}$Cs atom. It is an atom with a single valence
electron outside a closed-shell core. As the first step we carry out
finite-difference Dirac-Hartree-Fock calculations for Cs core. The
core orbitals are fed into the spline code where they are used to
compute the matrix elements of the $V_\mathrm{DHF}$ potential,
Eq.(\ref{Eq:VDHF}). As in Ref.~\cite{JohBluSap88} the numerical
accuracy is monitored by comparing the resulting quasi-spectrum with
the DHF energies from the finite-difference code.

\section{Numerical examples for Cs atom}
\label{Sec:Numerics} Here we provide numerical examples involving
both ND and DKB sets for Cs atom. In the two Sections immediately
following we compare the performance of the ND and the DKB sets. We
generate the quasi-spectrum using both ND and DKB sets and carry out
comparisons for single-particle energies and hyperfine integrals in
Section~\ref{Sec:EnHFI} and parity-nonconserving amplitude in
Section~\ref{Sec:PNC}.  Section~\ref{Sec:Spur} contains an analysis
of spurious states in the ND method. In Section~\ref{Sec:E2} we
analyze second-order energy corrections and discuss a strategy of
optimizing the size of the basis.

\subsection{Energies and hyperfine integrals}
\label{Sec:EnHFI} We compare  ND and DKB quasi-spectrums with
energies obtained using a finite-difference DHF code for the
low-lying valence states in Table~\ref{Tab:EnHFI}.  The ND and DKB
calculations were carried out using $N=40$ basis functions for
B-splines of order $k=7$. We used a cavity of radius $R=50$ bohr.
For the cavity of this size, only a few lowest-energy orbitals
remain relatively unperturbed by the cavity. From examining
Table~\ref{Tab:EnHFI},  it is clear that both ND and DKB sets have a
similar accuracy for energies.

In the second part of Table~\ref{Tab:EnHFI} we compare values of the
radial integrals entering matrix elements of the hyperfine
interaction due to the electric (EJ) and magnetic (MJ) multipolar
moments of a point-like nucleus
\begin{align*}
I_{EJ}\left(  n\kappa\right)    & =\int \frac{dr}{r^{J+1}}\left(
P_{n\kappa}^{2}\left(  r\right)  +Q_{n\kappa}^{2}\left(  r\right)
\right)
,\\
I_{MJ}\left(  n\kappa\right)    & =2\int \frac{dr}{r^{J+1}%
}P_{n\kappa}\left(  r\right)  Q_{n\kappa}\left(  r\right)  .
\end{align*}
The angular selection rules require $j\leq J/2$. We use identical
integration grid for all three cases (DHF,ND,DKB) listed in the
Table. The grid is sufficiently dense near the origin, it contains
about 100 points inside the nucleus. The numerical integration
excludes the first interval of the grid. From examining the Table we
find that the DKB set outperforms the ND basis. While the ND set
still recovers two-three significant figures for the magnetic-dipole
coupling, it produces wrong results for electric-quadrupole and
magnetic-octupole integrals. Certainly, the accuracy in the ND case
improves for a larger basis set, but larger basis sets come at an
additional computational cost. We  carried out similar comparisons
for matrix elements of the electric-dipole operator. As for the
energies, we find that both the ND and DKB sets perform with a
similar numerical accuracy.

As we have mentioned, $\left(  \Delta S^{\mathrm{spur}}\right)
_{\mathrm{ND}}$ variationally encourages the boundary condition
$P\left(  0\right) =0$. However, there is no such explicit
encouragement for $Q\left(  0\right)$. Here we qualitatively discuss
the observed properties of the radial components near the origin for
the Cs ND set. We see that, though their general behavior is to
approach zero, the large component wavefunctions often have small
improper inflections or oscillations near the origin. Small
component wavefunctions, on the other hand, often do not even
approach zero. These improper behaviors seem to be exemplified as we
look at states higher in the spectrum. As we have seen here, such
improper behavior of both the large and small component radial
functions near the origin prove detrimental for properties
accumulated near the nucleus. We conclude that while producing the
results of a similar quality for matrix elements accumulated far
from $r=0$, the DKB set provides a better approximation to the
atomic orbitals near the nucleus.

\begin{table*}[ht]
\begin{center}
\begin{tabular}[c]{lldddd}%
\hline\hline
 \multicolumn{1}{c}{State}&
 \multicolumn{1}{c}{Set}&
 \multicolumn{1}{c}{Energy}&
 \multicolumn{1}{c}{M1 HFI}&
 \multicolumn{1}{c}{E2 HFI}&
 \multicolumn{1}{c}{M3 HFI}   \\
 \hline
  $6s_{1/2}$
      &  FD & -0.1273680    &   1.114751[-1]   &                     &                 \\
      &  DKB & -0.1273674    &   1.114741[-1]   &                     &                 \\
      &  ND  & -0.1273682    &   1.121812[-1]   &                     &                 \\[1em]

  $7s_{1/2}$
     &  FD & -0.5518735[-1]  &   3.063077[-2]   &                     &                 \\
     &  DKB & -0.5518714[-1]  &   3.063069[-2]   &                     &                 \\
     &  ND  & -0.5518750[-1]  &   3.084164[-2]   &                     &                 \\[1em]

  $6p_{1/2}$
     &  FD & -0.8561589[-1]  &  -1.252026[-2]   &                     &                 \\
     &  DKB & -0.8561576[-1]  &  -1.252018[-2]   &                     &                 \\
     &  ND  & -0.8561616[-1]  &  -1.218362[-2]   &                     &                 \\[1em]

  $6p_{3/2}$
     &  FD & -0.8378548[-1]  &   4.649107[-3]   &       6.693978[-1]  &      8.725496[0]  \\
     &  DKB & -0.8378543[-1]  &   4.649117[-3]   &       6.694037[-1]  &      8.744759[0]  \\
     &  ND  & -0.8378538[-1]  &   4.649454[-3]   &       1.591405[+2]  &      1.924786[7]  \\[1em]

 $5d_{3/2}$
     &  FD & -0.6441964[-1]  &  -3.543808[-3]   &       1.702467[-1]  &     -8.950592[1]  \\
     &  DKB & -0.6441961[-1]  &  -3.543799[-3]   &       1.702736[-1]  &     -9.202786[1]  \\
     &  ND  & -0.6441970[-1]  &  -3.038702[-3]   &       1.643100[+5]  &      3.663198[10] \\[1em]

 $5d_{5/2}$
     &  FD & -0.6452977[-1]  &   2.257618[-3]   &       1.562954[-1]  &      1.938370[1]  \\
     &  DKB & -0.6452976[-1]  &   2.257616[-3]   &       1.562953[-1]  &      1.938705[1]  \\
     &  ND  & -0.6452969[-1]  &   2.257607[-3]   &       7.144243[+0]  &      7.235873[5]  \\
     \hline\hline
\end{tabular}
\end{center}
\caption{ Comparison of the energies and radial integrals of the
hyperfine interaction in the Dirac-Hartree-Fock approximation for
Cs. FD marks values produced by a finite-difference code. DKB and ND
values are generated with dual-kinetic-balance and Notre Dame
B-spline basis sets. In both cases we used $N=40$ basis functions
for B-splines of order $k=7$ in a cavity of $R=50$ bohr. Notation
$x[y]$ stands for $x \times 10^y$. \label{Tab:EnHFI}}
\end{table*}

\subsection{Parity-nonconserving amplitude}
\label{Sec:PNC} So far we examined properties of the individual
basis set orbitals with sufficiently low energies. The real power of
the finite basis set technique lies in approximating the entire
innumerable spectrum by a finite size quasi-spectrum. This is
important, for example, in computing sums over intermediate states
(Green's functions) in perturbation theory.

From the discussion of the preceding Section it is clear that the
difference in quality of the ND and DKB basis sets is expected to
become most apparent for the properties accumulated near the
nucleus. Here, as an illustrative example, we consider the
parity-nonconserving (PNC) amplitude for the $6S_{1/2} \rightarrow
7S_{1/2}$ transition in $^{133}\mathrm{Cs}$. This amplitude appears
in the second order of perturbation theory for the otherwise
forbidden dipole transition and it can be represented as a sum over
intermediate states $n p_{1/2}$
\begin{eqnarray}
E_\mathrm{PNC} &=& \sum_{n=2} \frac{\langle
7s_{1/2}|D|np_{1/2}\rangle  \langle np_{1/2} |H_{W}|6s_{1/2}\rangle
}{\varepsilon_{6s_{1/2}}-\varepsilon_{np_{1/2}}  }  \nonumber \\
&+ & \sum_{n=2} \frac{\langle 7s_{1/2}|H_{W}|np_{1/2}\rangle \langle
np_{1/2} |D|6s_{1/2}\rangle
}{\varepsilon_{6s_{1/2}}-\varepsilon_{np_{1/2}}} \, .~~
\label{Eq:EPNC}
\end{eqnarray}
Here $D$ and $H_{\rm W}$ are electric-dipole and weak interaction
(pseudo-scalar) operators, and $\varepsilon_{i}$ are atomic energy
levels. We will compute this expression in the single-particle
approximation. Specifically, the index $n$  runs over the entire DHF
quasi-spectrum for $p_{1/2}$ partial wave, including core orbitals.
The weak Hamiltonian reads
\begin{equation}
 H_{\rm W} = \frac{G_F}{\sqrt{8}} Q_{\rm W} \rho_{\mathrm nuc}(r) \gamma_5 \, ,
\label{Eq:Hw}
\end{equation}
where $G_F$ is the Fermi constant, $Q_{\rm W}$ is the weak charge,
$\gamma_5$ is the Dirac matrix (it mixes large and small
components), and $\rho_{\mathrm nuc}(r)$ is the neutron density
distribution. For consistency with the previous calculations the
$\rho_{\mathrm nuc}(r)$  is taken to be the proton Fermi
distribution of Ref.~\cite{BluJohSap92}. Notice that the matrix
elements of the weak interaction are accumulated entirely inside the
nucleus.

We evaluate the sum~(\ref{Eq:EPNC}) using the DKB and the ND sets
with $N=40$ basis functions of order $k=7$. The integration grids
are the same in both cases and include large number of points ($\sim
100$) inside the nucleus. The PNC amplitude is conventionally
expressed  in units of $10^{-11} i |e| a_0 (-Q_{\rm W}/N_n)$, where
$N_n=78$ is the number of neutrons in the nucleus of $^{133}$Cs. In
these units, the results are
\begin{eqnarray*}
E_\mathrm{PNC}^\mathrm{FD} &=&  - 0.740 ,\\
E_\mathrm{PNC}^\mathrm{DKB} &=& - 0.7395 \, (N=40,k=7),\\
E_\mathrm{PNC}^\mathrm{ND} &=& - 0.8546 \, (N=40,k=7).
\,
\end{eqnarray*}
The finite-difference value is taken from Ref.~\cite{BluJohSap92}.
Again we note that the DKB set offers an improved performance over
the Notre Dame set. Reaching the comparable accuracy in the ND
approximation requires a larger basis set. For example, $N=75, k=9$
ND set reproduces the DKB result for the PNC amplitude.

\begin{figure}[ht]
\begin{center}
\includegraphics[scale=0.8]{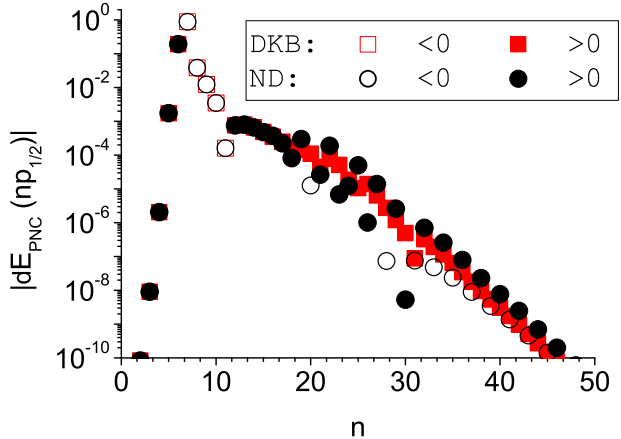}
\end{center}
\caption{(Color online) Individual contributions of the $p_{1/2}$ intermediate
states to the  PNC amplitude, Eq.(\ref{Eq:EPNC}),
as a function of the principal quantum number $n$. The units of the
PNC amplitude are $10^{-11} i |e| a_0 (-Q_{\rm W}/N_n)$. The results
from the DKB basis are represented by squares and those from
the ND basis by circles. Both sets contain $N=100$ basis orbitals and
use identical integration grids.
Due to the employed logarithmic scale, we plot the absolute values
of the contributions.
We use hollow symbols to indicate negative contributions and filled symbols
for marking positive contributions.
 } \label{Fig:PNCindividual}
\end{figure}

Further insights may be gained from examining individual
contributions of the intermediate states in the PNC amplitude. We
plot individual contributions of the intermediate state $np_{1/2}$
to the  PNC amplitude in Fig.~\ref{Fig:PNCindividual}.  Both terms
in Eq.(\ref{Eq:EPNC}) are included. We computed the data using
$N=100, k=11$ ND and DKB basis sets generated in a cavity of $R=50$
bohr.  From the plot, we observe that the dominant contribution
arises from the low-lying valence states. As $n$ increases, the
contributions become quickly suppressed (there is 10 orders of
magnitude suppression for $n\approx 50$). This is due to both
increased energy denominators and decreased density at the nucleus
for high $n$. Comparison between the basis sets reveals that their
contributions are identical until $n=17$. For higher  principal
quantum numbers, the DKB contributions monotonically decrease, while
ND contributions become irregular. Moreover, at $n=28$ the ND
contributions start to  flip signs with increasing $n$. Generally,
this oscillating pattern  would lead to a deterioration of the
numerical accuracy. We believe that the described irregularity is
again due to the aforementioned improper behavior of orbitals near
the nucleus, as the matrix elements of the weak Hamiltonian is
accumulated largely in this regime.



To summarize, the DKB set is numerically complete and is well suited
for carrying out practical summations over intermediate states in
perturbation theory. The comparison with the ND set shows that, at
least for the PNC amplitude, the ND convergence pattern becomes
affected by the inaccurate representation of the orbitals near the
nucleus (this is exemplified for states higher in the spectrum),
while the DKB set exhibits a monotonic convergence. Additionally,
the DKB set is devoid of spurious states, and therefore the
incidental inclusion of spurious states in summations over
intermediate states is of no concern.

\subsection{Analysis of spurious states in the ND method}
\label{Sec:Spur} Here we analyze the effect that the additional term
$\left( \Delta S^{\mathrm{spur}}\right) _{\mathrm{ND}}$ has on
spurious states which occur in the ND method. We start by taking

\begin{equation}
\Delta S^{\mathrm{spur}}=\left\{
\begin{array}[c]{cc}
x \frac{c}{2}P\left(  0\right)
^{2}-\frac{c}{2}P\left( 0\right) Q\left( 0\right),& \kappa<0\\
x c^{2}P\left(  0\right)
^{2}-\frac{c}{2}P\left( 0\right) Q\left( 0\right),& \kappa>0\\
\end{array}
\right.  ,
\end{equation}
with $x$ an adjustable parameter. For the case of $x = 1$ this is
equivalent to $\left(  \Delta S^{\mathrm{spur}}\right)
_{\mathrm{ND}}$. We mentioned previously that $\left(  \Delta
S^{\mathrm{spur}}\right) _{\mathrm{ND}}$ variationally encourages
the boundary condition $P\left(  0\right) =0$; this is also true for
any choice of $x$. The specific choice of $x$ effectively alters the
degree of such variational ``encouragement.''

\begin{figure}[t]
\begin{center}
\includegraphics[scale=1]{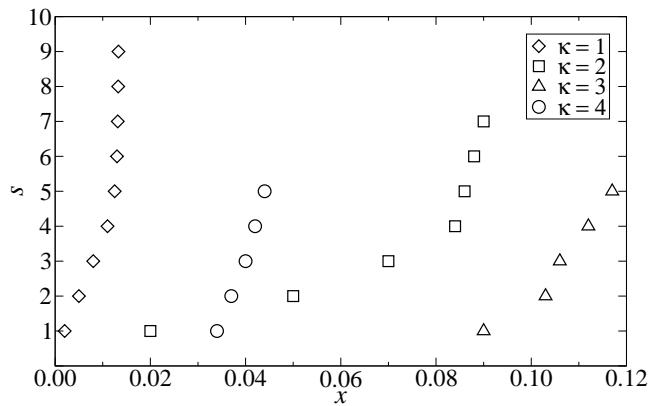}
\end{center}
\caption{Location of the spurious state in the spectrum of Cs for
$1\le\kappa\le4$ at various values of $x$. Here $s$ corresponds to
the location of the spurious state in the spectrum (e.g., $s=2$
corresponds to the spurious state appearing as the second lowest
energy state for that $\kappa$). We used a set with $N=40$ B-splines
of order $k=7$ confined to a cavity of radius $R = 50$ bohr.}
\label{Fig:SpurShift}
\end{figure}

We find that setting $x=0$ results in a single spurious state which
appears as the lowest energy eigenstate for each $\kappa>0$. By
subsequently increasing the value of $x$ towards $x=1$ we may then
deduce the effect that $\left( \Delta S^{\mathrm{spur}}\right)
_{\mathrm{ND}}$ has on these spurious states as well as the rest of
the spectrum. We find that small increases in $x$ from $x=0$ causes
the energy of the spurious state to increase, while the other states
remain essentially unaffected (except for the case of near
degeneracy with one of these states; this situation is discussed
below). As $x$ is increased from zero, we may watch as spurious
states for each $\kappa>0$ first appear as the lowest energy state,
then move up to the second lowest energy state, then to the third
lowest energy state, etcetera. Fig.~\ref{Fig:SpurShift} displays
this effect for the case of Cs.

\begin{figure}[t]
\begin{center}
\includegraphics[scale=1]{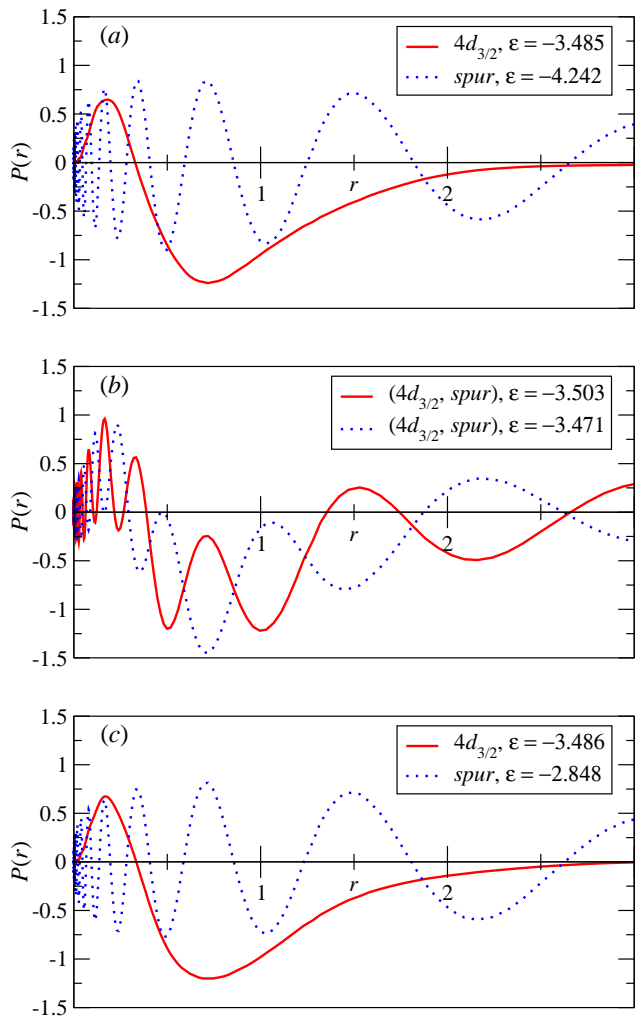}
\end{center}
\caption{(Color online) Behavior of the large component radial
wavefunction $P(r)$ of the $4d_{3/2}$ ($\kappa=2$) state of Cs when
the spurious state is ($a$) just below it in the spectrum
($x=0.052$), ($b$) nearly degenerate with it ($x=0.054$), and ($c$)
just above it in the spectrum ($x=0.056$). For the case ($b$), the
notation $(4d_{3/2},spur)$ is used to indicate that the states are
some linear combination of the two states and not easily defined as
one or the other. Here $\varepsilon$ refers to the energy in a.u.
with rest energy subtracted off. For reference, the ND set ($x=1$)
gives the lowest three $d_{3/2}$ states to have energies:
$3d_{3/2}$, $\varepsilon = -28.310$; $4d_{3/2}$, $\varepsilon =
-3.486$; $5d_{3/2}$, $\varepsilon = -0.064$. We used a set with
$N=40$ B-splines of order $k=7$ confined to a cavity of $R = 50$
bohr; note that these plots only extend to $r=3$ bohr, however.}
\label{Fig:Spur4d32}
\end{figure}

It is also interesting to analyze the effect of the spurious state
when its energy is nearly degenerate with another state in the
spectrum (we will refer to this other state as the ``genuine''
state). As the energies approach degeneracy by varying $x$, we
observe that the spurious state begins to mix in with the genuine
state. The first evidence for this is that the energy of the genuine
state starts to become affected by the presence of the spurious
state. Secondly, we see that the radial wavefunctions $P(r)$ and
$Q(r)$ of the genuine state begin to acquire non-physical ``bumps''
that oscillate in a way that corresponds to the rapid oscillations
of the respective spurious state radial functions. As the energy
becomes nearly degenerate, the two states mix to such a degree that
it is not even possible to define one as the ``genuine state'' and
the other as the ``spurious state.'' This effect is shown in
Fig.~\ref{Fig:Spur4d32}, where the genuine state is taken as the
4$d_{3/2}$ state of Cs. As $x$ is increased such that the spurious
state becomes embedded in the quasi-continuum part of the spectrum,
it mixes with multiple neighboring states, and at this point it
becomes difficult to track or even define the spurious state.

Presumably increasing $x$ to $x=1$ (the Notre Dame case) shifts the
spurious states all the way to the end of the spectrum. Evidence for
this is seen, for example, in the $d_{3/2}$ set used for
Fig.~\ref{Fig:Spur4d32} ($N=40$, $k=7$, $R = 50$ bohr). Here the
last few levels (excluding the very last level) have an energy
spacing on the order of $10^6$ a.u., whereas the energy spacing to
the final level is on the order of $10^9$ a.u. Furthermore,
increasing $x$ past $x = 1$ only has a substantial effect on the
energy of the final level (for $x = 10,000$ the energy spacing to
the final level is then on the order of $10^{13}$ a.u.).

Up to this point we have been exclusively considering cases with
$\kappa>0$, as these have been the only angular symmetries for which
spurious states have previously been known to occur. Now we shall
consider the effect that $\left( \Delta S^{\mathrm{spur}}\right)
_{\mathrm{ND}}$ has on the cases of $\kappa<0$. As with the
$\kappa>0$ cases, we see that increasing $x$ past $x = 1$ only has a
substantial effect on the energy of the final level. This is an
indication that a spurious state may actually lie at the end of the
spectrum for $\kappa<0$ angular symmetries as well. In fact, we find
that setting $x$ to a negative value significantly below $x = 0$
results in a single spurious state which appears as the lowest
energy eigenstate for each $\kappa<0$. By subsequently increasing
$x$ from this point, we may watch as each spurious state is shifted
towards the higher energy end of the spectrum, similar to what is
observed with $\kappa>0$ spurious states.

Now we return briefly to subject of the matrix elements $D_{1,n+1}$
and $D_{n+1,1}$ of the ND basis, which are suspected to contribute
to poor representation of orbitals near the nucleus. From
Eqs.~(\ref{Eq:Dij}) and (\ref{Eq:NDset}) we see that these include
the integral $\int \frac{1}{r} \left[ B_{i=1}^{\left( k\right)
}\left( r\right) \right]^2 dr$, which is infinite due to the
non-vanishing property of the first B-spline at $r=0$. Numerically
this integral is evaluated by Gaussian quadrature, producing finite
values. We observe that for the ($N=40$, $k=7$, $R = 50$ bohr) Cs
set, increasing the numerical value of this integral to 20 times its
Gaussian quadrature value results in the reappearance of spurious
states as the lowest energy eigenstates for each $\kappa>0$.
Simultaneously, the energy of highest energy eigenstate for each
$\kappa<0$ is increased substantially. Evidently the capability of
$\left( \Delta S^{\mathrm{spur}}\right) _{\mathrm{ND}}$ to shift the
spurious state to the end of the spectrum for $\kappa>0$ angular
symmetries is reliant upon the inaccurate numerical evaluation of
this (theoretically) infinite-value integral.

The claim made in this section of observing spurious states for
$\kappa<0$ angular symmetries may seem surprising at first.
\citet{ShaTupYer04} have proved that an independent expansion of
large and small radial components with a finite set of basis
functions leads to a single spurious state for $\kappa>0$ angular
symmetries. This proof assumes the basis functions to vanish at the
origin, and the result of this proof is consistent with experience
when such basis functions are employed. Arguably, connection is lost
immediately with this proof because the ND set includes the first
B-spline, which does not vanish at the origin. As we have seen here,
the ND method depends on numerical inaccuracies in evaluating
infinite-value integrals in order to manipulate spurious states
arising in $\kappa>0$ cases. Because the ND method also includes the
first B-spline for $\kappa<0$ angular symmetries (and hence similar
numerical inaccuracies), we would have no reason to discount the
possibility of spurious states from occurring in these cases as
well.


\subsection{Optimizing the basis set: second-order energy correction}
\label{Sec:E2} In the preceding sections we established that the DKB
sets are more robust than the ND bases. For a comparable size of the
basis set the DKB basis exhibits better numerical accuracy for
properties accumulated at small radii. Or, we may say that for a
fixed numerical accuracy, the DKB basis may contain smaller number
of basis functions. In this section we investigate a related
question: what the smallest possible basis set is for a given
numerical accuracy. Keeping the set as small as possible speeds up
many-body calculations that usually require multiple summations over
intermediate states. Also, smaller basis sets reduce storage
requirements for  expansion coefficients in all-order techniques
such as configuration interaction or coupled-cluster methods.

Quantifying the numerical accuracy requires choosing some metric, which
characterizes deviation of the selected property for a given basis from its
exact value.  Apparently, one should select  the ``metric'' so
that it can be easily computed and is related to the relevant atomic properties.
As an example of an optimization measure, here we
choose the second-order correction to the energy of a valence electron,
$E_{v}^{\left(  2\right)  }$.

In the frozen-core DHF approximation $E_{v}^{\left(  2\right)  }$ is the leading many-body correction to
the energy.   It is given in terms of the Coulomb integrals
$g_{ijkl} = \int d1 d2 u_i^\dagger(1) u_j^\dagger(2) \left(1/r_{12} \right) u_k(1) u_l(2)$
and single particle energies $\varepsilon_{i}$
as (see, e.g., Ref.~\cite{Joh07book}) ,
\begin{equation}
E_{v}^{\left(  2\right)  }=\sum_{abn}\frac{\tilde{g}_{abvn}g_{vnab}%
}{\varepsilon_{v}+\varepsilon_{n}-\varepsilon_{a}-\varepsilon_{b}}-\sum
_{mna}\frac{\tilde{g}_{vamn}g_{mnva}}{\varepsilon_{m}+\varepsilon
_{n}-\varepsilon_{v}-\varepsilon_{a}} \, ,
\label{Eq:E2v}
\end{equation}
where $\tilde{g}_{ijkl} =   {g}_{ijkl} -  {g}_{ijlk}$ is the antisymmetrized Coulomb integral.
The summations are carried out over core orbitals (labels $a$ and $b$) and virtual (non-core)
orbitals (labels $m$ and $n$). Each summation implies summing over principal quantum
numbers, angular numbers $\kappa$, and magnetic quantum numbers. The summation over
magnetic quantum numbers
may be carried out analytically and we are left with summations over radial functions.

We define a contribution of an individual partial wave $\ell$,
$\delta E^{(2)}_{v} (\ell)$, as the difference
 $\delta E^{(2)}_{v} (\ell) =E^{(2)}_{v} (\ell) -E^{(2)}_{v} (\ell-1) $,
where $E^{(2)}_{v} (\ell)$ stands for truncated Eq.~(\ref{Eq:E2v});
it includes summations over intermediate states (both core and
virtual) with the orbital angular momentum up to $\ell$. Since the
calculations necessarily involve Coulomb integrals between orbitals
of different angular momenta, the numerical error in $\delta
E^{(2)}_{v} (\ell)$ is affected by the accuracy of representation of
all partial waves up to $\ell$.

First, in Table~\ref{Tab:EnSecond}, we present results for
a large set of $N=100$ basis functions for each partial wave. We use a sufficiently large cavity of $R=50$ bohr in this
calculation. These results will serve as a benchmark for comparisons
with the less complete (optimized) sets. We observe that the dominant (60\%) contribution
comes from $d$-waves, $\ell=2$. Qualitatively, the second-order energy correction
may be described as core-polarization effect.  The outer $3d$-orbitals of the core
are relatively ``soft'' and are easily polarizable. Since the core does not contain $f$ and
higher partial waves, after peaking  at $\ell=2$, the partial-wave contributions
become suppressed as $\ell$ increases.  Qualitatively, this suppression arises due to increased
centrifugal repulsion of higher partial waves and  associated
smaller overlaps with core orbitals in the Coulomb integrals of Eq.~(\ref{Eq:E2v}).
For example, $\ell=6$ contributes only 0.6\% of the total value.

Now we would like to minimize the size of the set
by choosing  a different number of radial basis functions $N_\kappa$
for  different angular symmetries $\kappa$. To preserve a numerical
balance between the fine-structure components (for example, this may be
important while recovering non-relativistic limit) we keep the same
number of functions for a given orbital angular momentum $\ell$, e.g., $N_{p_{1/2}} = N_{p_{3/2}}$.

In Table~\ref{Tab:EnSecond},
we present an example of an optimized basis (marked as ``Small set'').
We also list resulting numerical errors for each partial wave by comparing
$\delta E^{(2)}_{6s} (\ell)$ with the result from the ``Large set'' calculations.
We see that while the basis-set error in higher partial waves is
as large as 40\%, this hardly makes any influence on the total value
of the correlation energy, because of contributions of higher $\ell$ are suppressed.
The total value of the correlation correction differs by about 1\% from it's
saturated value. Considering that the correlation contribution to the energy
is about 10\% for the $6s$ state, the less complete set would introduce only 0.1\% error
for the total ionization energy.

So far we discussed the correlation energy correction for the
$6s_{1/2}$ state. The optimized set remains sufficiently robust for
other low-lying states as well. We have carried out a comparison
similar to Table~\ref{Tab:EnSecond} for $6p_{1/2}$, $6p_{3/2}$,
$5d_{3/2}$, and $5d_{5/2}$ states. In all these cases the difference
between the total  $E^{(2)}_{v}$ values computed with the``Large''
and ``Small'' sets is about 0.5\%. In practical calculations one is
often required to reproduce a number of properties with the same
set. The atomic properties may be quite dissimilar-- like
hyperfine-structure interactions accumulated near the nucleus and
dipole matrix elements determined by the valence region. Apparently,
one has to carry out a similar low-order MBPT analysis for the
relevant quantities to verify the suitability of the optimized set.

From the computational point of view, using the optimized sets
speeds up the numerical evaluations. In our illustrative example,
the ``Large set'' contains $N_t=(1+2\times6) \times 100 = 1300$
orbitals, while the ``Small set'' is about four times smaller
($N_t=285$ orbitals).  The resulting speed-up is sizable: our
computations of Eq.~(\ref{Eq:E2v}) for the $6s_{1/2}$ state
($\ell=6$) are about 14 times faster with the optimized set, as
expected due to $N_t^2$ scaling of the number of contribution in the
most computationally demanding second term  of  Eq.(\ref{Eq:E2v}).
Similar scaling should hold for storage of expansion coefficients in
all-order methods, e.g., for storing triple
excitations~\cite{DerPor07} one expects $N_t^3$ scaling of the
storage size. Usually higher partial waves produce larger number of
angular channels and the scaling should be even steeper than
$N_t^3$.
Further speed-up in MBPT summations and reduction in storage size may be attained by
skipping a few basis set functions at the upper end of the quasispectrum. Such additional
truncation of the spectrum  becomes apparent from Fig.~\ref{Fig:PNCindividual},
where contributions to the PNC amplitude for the upper three-quarters ($n \gtrsim 25$ out of $N=100$) of
the quasi-spectrum affect the total value below $10^{-4}$ level of accuracy.

\begin{table}[ht]
\begin{center}
\begin{tabular}[c]{lclclr}%
\hline\hline
 \multicolumn{1}{c}{$\ell$}&
 \multicolumn{2}{c}{Large set}&
 \multicolumn{2}{c}{Small set}&
 \multicolumn{1}{c}{Error}\\
 \hline
   & $(N,k)$  &  \multicolumn{1}{r}{$\delta E^{(2)}_{6s} (\ell)$}
   & $(N,k)$  &  \multicolumn{1}{r}{$\delta E^{(2)}_{6s} (\ell)$}    \\
 0 & (100,11) &  -0.0000130 &(35,7)&-0.0000122 & 6\%   \\
 1 & (100,11) &  -0.0020027 &(35,7)&-0.0019936 & 0.5\% \\
 2 & (100,11) &  -0.0105623 &(30,5)&-0.0105373 & 0.2\% \\
 3 & (100,11) &  -0.0039347 &(25,4)&-0.0039095 & 0.6\% \\
 4 & (100,11) &  -0.0007563 &(15,4)&-0.0007269 & 4\%   \\
 5 & (100,11) &  -0.0002737 &(10,4)&-0.0002272 & 20\%  \\
 6 & (100,11) &  -0.0001182 &(10,4)&-0.0000844 & 40\%  \\
\hline
\multicolumn{1}{l}{Total } &
              &   -0.0176609 &     & -0.0174912 & 1\% \\
 \hline\hline
\end{tabular}
\end{center}
\caption{  Contribution of individual partial waves $\ell$ to the
second-order energy correction for the ground $6S_{1/2}$ state of Cs
atom. We use the DKB basis set of $N$ basis functions constructed
from a subset of B-splines of order $k$ (label  $(N,k)$).
Calculations are carried out in a cavity of $R=50$ bohr for two
basis sets  ``Large'' and ``Small''. Numerical integration grids are
identical for both sets. The column marked ``Error'' refers to a
relative error in a given partial-wave contribution, $\delta
E^{(2)}_{6s} (\ell)$, caused by switching from the ``Large'' to the
``Small'' basis set. \label{Tab:EnSecond}}
\end{table}

\section{Conclusion}
Calculations of certain atomic properties, such as parity-violating
amplitudes, transition polarizabilities between hyperfine levels,
and many-body corrections to hyperfine interactions require accurate
representation of atomic orbitals at both small and intermediate
electron-nucleus distances. Solving the many-body problem in high
orders of perturbation theory additionally calls for an efficient
representation in terms of the basis sets. The dual-kinetic-basis
set is shown here to adequately meet both these demands.

Previously the DKB basis was applied to systems with a single
electron in a Coulomb field, i.e., hydrogen-like
ions\cite{ShaTupYer04,Iga06}. Here we extended the DKB method to
many-electron systems by generating the single-particle
quasi-spectrum of the Dirac-Hartree-Fock potential. Several
numerical example for Cs atom were presented. We showed that the DKB
method outperforms the widely-employed Notre Dame B-spline
method~\citet{JohBluSap88} for problems involving matrix elements
accumulated at small distances.

In addition, we presented a strategy for optimizing the size of the
basis sets by choosing progressively smaller number of basis
functions for increasingly higher partial waves. This strategy
exploits suppression of contributions of high partial waves to
typical many-body correlation corrections.

\section{Acknowledgments}
We would like to thank Walter Johnson, Charlotte Froese-Fischer, and
Vladimir Shabaev for discussions. The developed dual-kinetic-balance
code for the Dirac-Hartree-Fock problem was partially based on codes
by the Notre Dame group. This work was supported in part by the
National Science Foundation.

\bibliographystyle{apsrev}

\end{document}